\documentclass[10pt,twocolumn,letterpaper]{article}
\usepackage{hyperref}

\makeatletter
\newcommand{\printfnsymbol}[1]{%
  \textsuperscript{\@fnsymbol{#1}}%
}
\makeatother

\usepackage{wacv}
\usepackage{times}
\usepackage{graphicx}
\usepackage{amsmath}
\usepackage{amssymb}
\usepackage{float}
\usepackage{epstopdf}
\usepackage{indentfirst}

\usepackage{subcaption}
\usepackage{multirow}
\usepackage{enumitem}
\usepackage{wrapfig}
\newcommand\width{1.75in}
\newcommand\nwidth{\textwidth}

\wacvfinalcopy 

\def\wacvPaperID{268} 

\ifwacvfinal\fi
\begin{document}

\title{CompressNet: Generative Compression at Extremely Low Bitrates}

\author{Shubham Dash\thanks{Equal contribution.} , Giridharan Kumaravelu\printfnsymbol{1}, Vijayakrishna Naganoor\printfnsymbol{1}, \\ Suraj Kiran Raman\printfnsymbol{1}, Aditya Ramesh\printfnsymbol{1}, Honglak Lee 
 \\
University of Michigan, Ann Arbor\\
}


\maketitle
\ifwacvfinal\thispagestyle{empty}\fi

\begin{figure*}[h]
\begin{center}
  \includegraphics[width=0.85\textwidth, height=7cm]{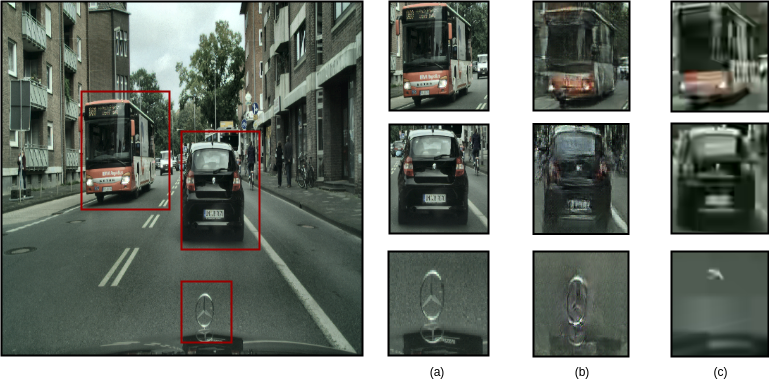}
  \end{center}
  \caption{Original Image, a) Original image patches, b) SAE-SPN (ours) image patches, and c) BPG image patches \cite{BPG}}.
\end{figure*}
 
\begin{abstract}
Compressing images at extremely low bitrates ($<$ 0.1 bpp) has always been a challenging task since the quality of reconstruction significantly reduces due to the strong imposed constraint on the number of bits allocated for the compressed data. With the increasing need to transfer large amounts of images with limited bandwidth, compressing images to very low sizes is a crucial task. However, the existing methods are not effective at extremely low bitrates. To address this need, we propose a novel network called CompressNet which augments a Stacked Autoencoder with a Switch Prediction Network (SAE-SPN). This helps in the reconstruction of visually pleasing images at these low bitrates ($<$ 0.1 bpp). We benchmark the performance of our proposed method on the Cityscapes dataset, evaluating over different metrics at extremely low bitrates to show that our method outperforms the other state-of-the-art. In particular, at a bitrate of 0.07, CompressNet achieves $22\%$ lower Perceptual Loss and $55\%$ lower Frechet Inception Distance (FID) compared to the deep learning SOTA methods. 
\end{abstract}


\section{Introduction}

With the exponential growth of visual data-transfer, effective compression to extremely small scales is of paramount significance. In the case of images, classical image compression techniques, such as JPEG \cite{JPEG}, WebP \cite{WebP}, and BPG \cite{BPG} fail to generate high quality reconstructions at low bitrates. However, lossy compression techniques using generative compression \cite{abs-1804-02958}, \cite{rippel2017real}, and \cite{GC} show promise in the reconstruction of aesthetically pleasing images at similar operating conditions.

Any lossy image compression scheme can be formulated as a rate-distortion optimization problem. In this framework with an autoencoder setup, an analysis transformation, $f$ : $\mathbb{R}^N$ $\rightarrow$ $\mathbb{R}^M$, maps the input data $x$ to a vector $z$ in latent space, and a synthesis transform, $g$ : $\mathbb{R}^M$ $\rightarrow$ $\mathbb{R}^N$ , transforms $z$ back into the image space.

A majority of the existing compression systems are optimized for distortion metrics, such as peak signal-to-noise ratio (PSNR) or different variants of structural similarity (SSIM) (Wang \textit{et al.}, 2003). Traditionally, the focus has been put on building hand-crafted codecs (encoder-decoder pairs for compression tasks) by making strong assumptions, such as the codec applying linear transform, as has been done with JPEG and JPEG2000. This assumption has an inherent problem as it is inaccurate to assume that a linear codec can generalize to compress a wide variety of natural images.

For extremely low bitrates, traditional metrics lose their relevance as they favor pixel-wise preservation of local structure over preserving texture and global structure. Recent works by Patel \textit{et.al} in \cite{reviewer1a}, \cite{reviewer1b} and Blau \textit{et.al} in \cite{perceptual_quality} indicate the need for more accurate perceptual metrics to evaluate the visual quality of the images, rather than evaluating the structural similarity as captured by the traditional metrics. For a compression task, the reconstructions require high perceptual quality and closely resembled the original image. Training a system with adversarial losses in this scenario produces more accurate results as it enables an improved understanding of the global structure of the image.  We integrate a Generative Adversarial Network (GAN) setup along with the autoencoder for achieving this task.

Stacked-autoencoders that incorporate layer-wise loss for learning latent dimensions for supervised tasks have been proven to enhance reconstruction quality for image compression tasks \cite{SAE_all} over traditional autoencoders. We incorporate a similar idea in CompressNet for enhancing the reconstruction quality at extremely low bitrates. In addition to stacked autoencoders, Stacked What-Where Autoencoder(SWWAE) \cite{ZhaoMGL15} models suggest the use of pooling switch information for improved data-reconstruction across the encoder-decoder architectures. However, incorparating the pooling switch information increases data overhead making it infeasible for image-compression tasks at extremely low bitrates. In order to incorporate SWWAE models for compression tasks with no additional data-overhead, we propose a network to predict pooling switches and use it along with the SAE-All architecture. This allows us to operate at extremely low bitrates with just minimal computational overhead demonstrating comparable performance to SWWAE, which has proven to perform appreciably well for compression tasks.

The main contributions of the paper for image compression at extreme bit rates include, \begin{itemize}
    \setlength\itemsep{0em}
    \item Stacked Autoencoder (SAE) and Stacked What-Where Autoencoder (SWWAE) based architectures with layer-wise loss. 
    \item Stacked Autoencoder with Switch Prediction Network (SAE-SPN) with layer-wise loss.
    \item Benchmarking the proposed architectures which show 22 \% lower Perceptual Loss and 55 \% lower Frechet Inception Distance (FID) compared to the deep learning SOTA methods at a bitrate of 0.07.
\end{itemize}

\section{Related Work}
The classical approach to the compression theory, which is mathematically formulated by Shannon's theory of communication \cite{Shannon}, provides the fundamental basis for the coding theory. Classical methods leverage explicit probabilistic modelling and feature extractions, effectively engineered for the task of image-compression \cite{vk}, JPEG \cite{JPEG} and BPG \cite{BPG}. The application of deep learning for image compression has recently emerged as an active area of research. Incorporating autoencoder models into compression frameworks remains to be one of the most popular approaches amongst the deep learning techniques. Theis \textit{et al.} \cite{lossy}, Balle \textit{et al.}\cite{Balle},  Toderici \textit{et al.}\cite{DBLP}, Lee \textit{et al. }\cite{adaptive}, and Minnen \textit{et al.} \cite{David}  have employed DNN architectures successfully for the task of image compression. Along with the autoencoders, GANs \cite{GoodfellowPMXWOCB14} have also been considered as an alternative to the more traditional approaches, such as JPEG \cite{JPEG} and BPG \cite{BPG}. GANs tend to produce more aesthetically pleasing and accurate reconstructions. In this section, we specifically review image-compression frameworks that incorporate autoencoders and GANs.

\paragraph{}
An autoencoder is a neural network that learns to reconstruct the input. It contains a latent layer describing a code used to represent the input in order to reconstruct it. Autoencoders are constrained to be incapable of optimizing directly due to the inherent non-differentiability of the compression loss. Mean-squared loss is generally used to measure the degree of distortion between the original and reconstructed images and is used to optimize the encoder-decoder network. Theis et al. \cite{lossy} proposed a method to overcome this problem and have shown that minimal changes to the loss are sufficient to train deep autoencoders that are at par with JPEG 2000 in terms of the degree of compression making it suitable for compressing high-resolution images.
Alexendre \textit{et al.} (2018) \cite{DBLP:journals/corr/abs-1902-07385}  proposed using autoencoders along with residual blocks and skip connections to achieve lossy compression at low bitrates $(\backsim 0.15 bpp)$. However, this approach suffers at extremely low bitrates $(\leq 0.1 bpp)$ because it optimizes for MS-SSIM, which emphasizes the pixel -level preservation of an image, leading to blurry reconstructions.

\begin{figure*}[h]
\begin{center}
  \includegraphics[width=0.85\textwidth]{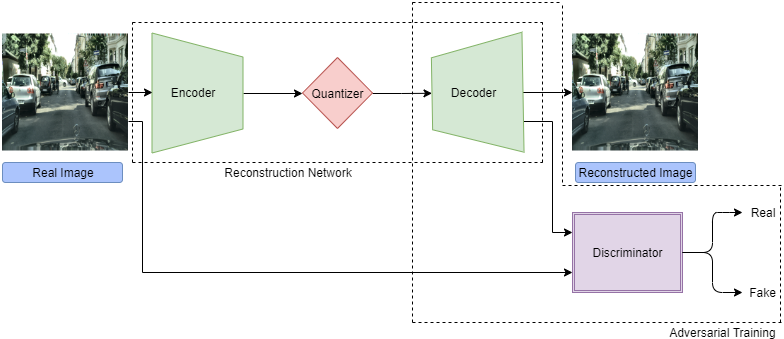}
  \end{center}
  \caption{CompressNet Architecture}
  \label{fig:2}
\end{figure*}

\paragraph{}
GANs have been used for learning intractable distributions in an unsupervised manner. At extemely low bit-rates, compression networks based on reconstruction losses prove to be ineffective as they learn unimodal approximations of the real distribution. Fingscheidt \textit{et al.} \cite{SS_GC} showed using GAN architectures that traditional compression algorithms and techniques, which use reconstruction loss to optimize for image compression, lead to high PSNR and MS-SSIM, but this does not guarantee accurate perception functions, in this case, semantic segmentation. This necessitates the usage of GAN to capture the global structure and context of the image, enabling extreme learned compression. 
Given a data set X, GANs approximate its (unknown) distribution $p_x$ through a generator G(z) that tries to map samples z from a fixed prior distribution $p_z$ to the distribution $p_x$. This helps the generator to reconstruct the sharper images.
The generator G is trained in parallel with a discriminator D by searching for a saddle point of a mini-max objective. Taking into consideration the reconstruction error and adding the corresponding loss term $\mathbb{E}[d(x,g(D(G(z)))]$, which refers to the Vanilla GAN loss proposed in \cite{GoodfellowPMXWOCB14} to the total loss, 
the objective function is now:
\begin{equation}
\begin{split}
\min_G \max_D \mathbb{E}[f(D(x))] &+ \mathbb{E}[g(D(G(z)))] \\&+ \lambda \mathbb{E}[d(x,g(D(G(z)))] \text{,}
\end{split}
\end{equation}

which implies minimising $\mathbb{L}_{GAN}$ over G.
\begin{equation}
\begin{split}
\mathbb{L}_{GAN} := \max_D \mathbb{E}[f(D(x)] &+ \mathbb{E}[g(D(G(z))) \\ &+ \lambda \mathbb{E}[d(x,g(D(G(z)))] \text{,}
\end{split} 
\end{equation}
In this work, we use $f(y)$ = $\log(y)$ and $g(y)$ = $\log(1-y)$ used in Vanilla GAN proposed by Goodfellow et al.\cite{GoodfellowPMXWOCB14} which implies that we are finding the G minimizing the JS Divergence between the distribution of $x$ and $G(z)$. 

The effectiveness of Generative Compression is attributed to the decoder being adversarially trained with a "paired" discriminator, similar to how a GAN is trained. This allows the decoder to learn the real distribution of the data and helps it generate visually pleasing reconstructions from a compressed latent representation.

Recently the image and video compression research community has increasingly shown a strong penchant towards the usage of GANs. The early work by Santurkar \textit{et al.} \cite{GC} employs a GAN framework for image compression. Although they efficiently justify the potential of GANs, the work is more oriented towards representation learning on thumbnail images and not full resolution images. Rippel \textit{et al.} in \cite{rippel2017real} proposed an adversarial framework for compression that was primarily intended towards minimizing artifacts using an adversarial loss term and focuses on generating visually pleasing reconstructions. Presently, Agustsson \textit{et al.} in \cite{abs-1804-02958} propose two networks for general and selective compression using conditional GANs which is the current state-of-the-art compression standard at extremely low bitrates. It has been inferred in \cite{abs-1804-02958} that the usage of conditional GANs is more pronounced in the case of selective object-based compression over general compression, which is the current objective. 

\begin{figure*}
\hspace{0.2 in}
\subcaptionbox{SAE/SWWAE Architecture\label{fig:3a}}
{\includegraphics[height=5.5cm, angle=0]{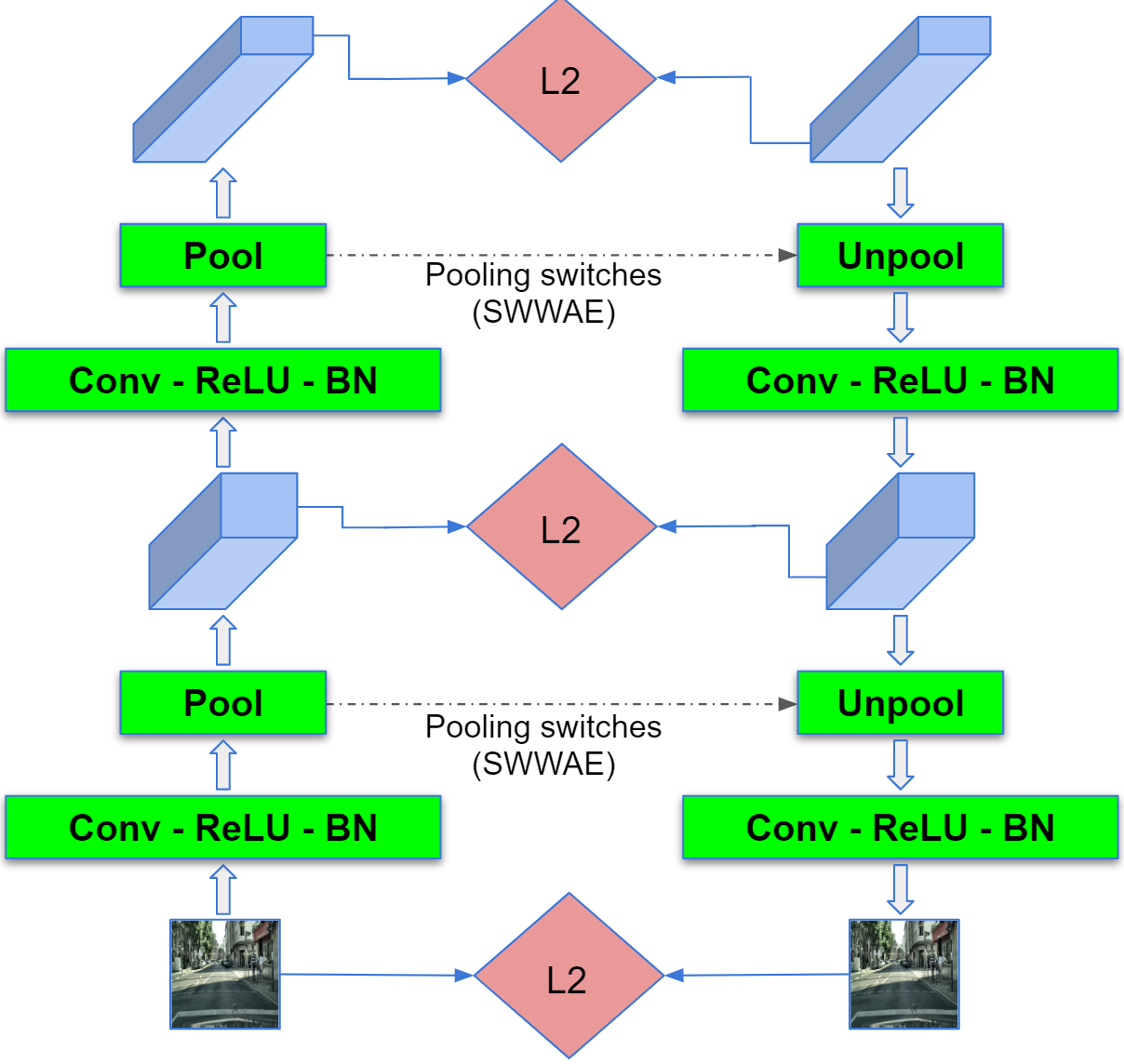}
}\hspace{0.2 in}\subcaptionbox{SAE-SPN Architecture\label{fig:3b}}
{\includegraphics[height=5.5cm, angle=0]{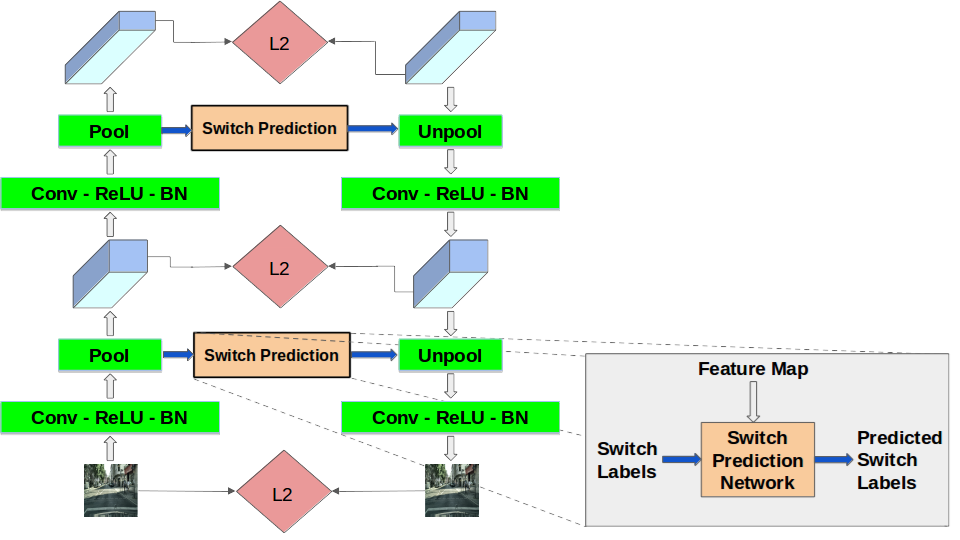}}

\caption{Architecture Description}
\end{figure*}

\section{Method}
The architecture for extreme learned compression (Figure \ref{fig:2}) has been inspired by the recent work proposed by Agustsson \textit{et al.} in \cite{abs-1804-02958}, specifically, the architecture of the encoder E and the generator G proposed in  Wang \textit{et al.} \cite{DBLP:journals/corr/abs-1711-11585}. The detailed explanation of both these architectures is explained below with the help of the diagrams.


The encoder takes in the image and converts it into a compressed feature space which is then passed through the Quantizer. Quantizer assigns a quantized value to each value in the compressed feature space based on the nearest quantization level, to obtain $\hat{w}$ a compressed and quantized representation. This forms the latent dimension from which the Decoder learns to reconstruct the original image back. This latent representation is then passed into the Generator or Decoder G, which produces the reconstructed image $\hat{x}$. The discriminator D is used for adversarial training which takes in this reconstructed image and the actual image and predicts whether the given image is real or reconstructed. The discriminator follows the PatchGAN architecture \cite{isola2017image}. A PatchGAN discriminator maps a 512 $\times$ 512 array to a $N \times N$ array of outputs X, where each $X_{ij}$ signifies whether the $patch_{ij}$ in the image is real or fake.

\subsection{Approach}
In order to improve the quality of reconstructions compared to existing to generative compression methods, we adopt three approaches as mentioned before. Each approach aims to modify the autoencoder setup in our model. These approaches are as follows:
\begin{figure*}[!h]
\begin{subfigure}[b]{0.25 \linewidth}
\centering\includegraphics[width = \width]{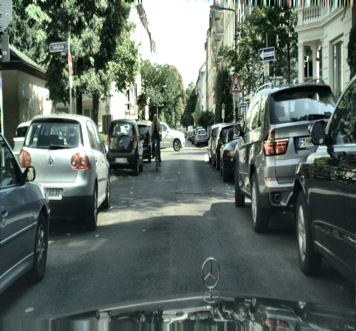}
\caption{Original Image \vspace{5pt}}
\end{subfigure}\hfill
\begin{subfigure}[b]{0.25 \linewidth}
\includegraphics[width = \width]{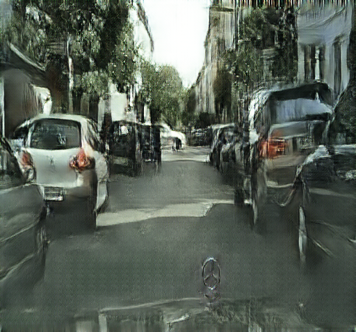}
\caption{SAE-All (ours) @ 0.073 bpp \vspace{5pt}}
\end{subfigure}\hfill
\begin{subfigure}[b]{0.25 \linewidth}
\includegraphics[width = \width]{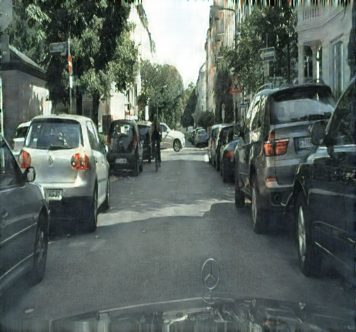}
\caption{SWWAE (ours) @ 56.31 bpp \vspace{5pt}}
\end{subfigure}\hfill
\vfill
\begin{subfigure}[b]{0.25 \linewidth}
\includegraphics[width = \width]{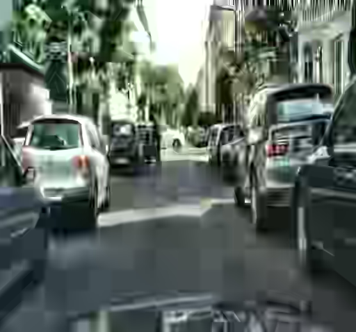}
\caption{BPG \cite{BPG} @ 0.0726 bpp}
\end{subfigure}\hfill
\begin{subfigure}[b]{0.25 \linewidth}
\includegraphics[width = \width]{saeallhigh.png}
\caption{SAE-SPN (ours) @ 0.073 bpp}
\end{subfigure}\hfill
\begin{subfigure}[b]{0.25 \linewidth}
\includegraphics[width = \width]{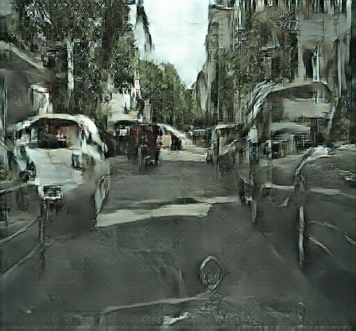}
\caption{SAE-All (ours) @ 0.036 bpp}
\end{subfigure}\hfill
\caption{Visual benchmarking of our proposed models with classical state-of-the-art method BPG\cite{BPG} }
\label{fig:ovr}
\end{figure*}

\subsubsection{Stacked Autoencoders}

In this model, as shown in Figure \ref{fig:3a}, we compute layer-wise loss, add it to the final objective function, and then optimize the entire network jointly. The layer-wise loss is calculated by taking the $L^2$ norm between the layer responses after every MaxPool-MaxUnpool operation in the encoder-decoder architecture. This ensures that resulting reconstructions are as similar as possible, even in the feature space, and allows the encoded information to be propagated deeper into the encoding network, with minimal loss of information.

\subsubsection{Stacked What-Where Autoencoders}

Including the pooling switch information infuses the decoder network with missing information. As a result, during the reconstruction phase, the individual activations are placed where the maximum activation was observed during max-pooling in the encoding stage. However, this extra information causes the switch information to transmit from encoder to decoder. This increases the information overhead during compression and makes extreme compression infeasible since information transmission has to be minimized while keeping the reconstructions sharp.

\subsubsection{Stacked Autoencoders with Switch Prediction Network}

Even though the SWWAE architecture provides visually pleasing reconstructions the information overhead must be eliminated in order to be suitable for extreme compression. Incorporating pooling switch information is beneficial since the quality of the reconstructions obtained is significantly sharper. To retain the performance of the network in terms of perceptual quality along with traditional metrics like PSNR and $FSIM_{C}$, we predict the pooling switches using an auxiliary Switch Prediction Network (SPN) (Figure \ref{fig:3b}) which is a convolutional neural network with a sigmoid activation function in the output. We assume a $2 \times 2$ max-pool operation in the encoder, which maps a four element patch to a single value. 0 represents the top-left value in the $2 \times 2$ patch, 1 represents the top right, 2 represents the bottom left value, and 3 represents the bottom right in the patch.
For our experiments, a $3 \times 3$ kernel regresses values in the range of 0-1, then classes for predicting the max-pooling location are assigned as class 0 for values between 0-0.25, class 1 for values between 0.25-0.5, class 2 for values between 0.5-0.75 and class 3 for values between 0.75-1. The functioning of this variant remains the same as SWWAE with the switch prediction network deployed in the decoder of the overall architecture. \\
Figures \ref{fig:3a} and \ref{fig:3b} explain how both architectures are used for extreme compression. In both figures, the left side represents the encoder and the right side represents the decoder. a) signifies the SAE and SWWAE architecture which is designed similarly with layerwise $L_{2}$ loss between each layer of encoder and decoder.
Part b) of the figure describes SAE-SPN architecture. The salient difference is that instead of passing the pooling switch information like in SWWAE, we predict pooling switch based on decoder response. SWWAE uses pooling switch information to reconstruct the pixels at the exact location from where the max activation had taken place. Intuitively since we are predicting the pooling switches in SAE-SPN and no pooling switch information in SAE-All, the reconstructed pixels might not be always in the correct location of max activations giving a slightly inferior performance compared to SWWAE but has no information overhead. This makes it very feasible to be be used in extreme level compression.

\begin{figure*}[!h]
\begin{subfigure}[b]{0.3 \linewidth}
\includegraphics[width = \nwidth]{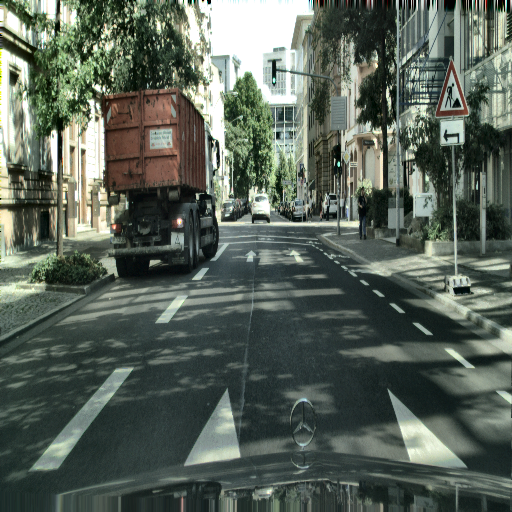}
\caption{Original Image}
\end{subfigure}\hfill
\begin{subfigure}[b]{0.3 \linewidth}
\includegraphics[width = \nwidth]{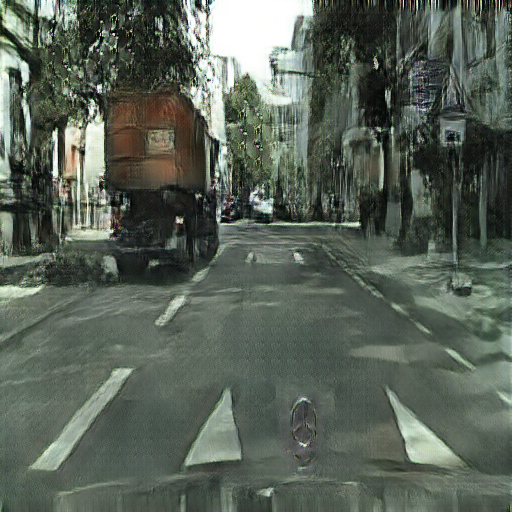}
\caption{SAE-All (ours) @ 0.073 bpp}
\end{subfigure}\hfill
\begin{subfigure}[b]{0.3 \linewidth}
\includegraphics[width = \nwidth]{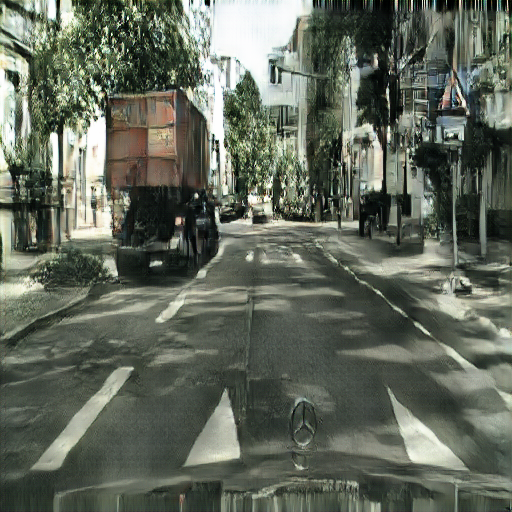}
\caption{SAE-SPN (ours) @ 0.073 bpp}
\end{subfigure}\hfill
\caption{Comparison of CompressNet with SAE-All method; CompressNet reconstructs with greater detail}
\end{figure*}

\subsection{Loss Function}
The loss function used to optimize the entire pipeline consists of,

\begin{itemize}
    \item Vanilla GAN loss function to optimize the generator and the discriminator, $L_{GAN}$
    \item Mean Squared Loss (MSE) to force the output reconstruction to be similar to the input image, $L_{MSE}$
    \item Perceptual Loss component to take care of the textural and feature similarity between the input and output images by minimizing the $L^2$ distance between the response from the 4$^{th}$ convolutional layer of a pre-trained Alexnet, $L_{perceptual}$
    \item SAE layer-wise loss the $L^2$ norm between the layer responses after every MaxPool - MaxUnpool operation in encoder-decoder architecture, $L_{SAE}$
\end{itemize}

\vspace{-1mm}

\begin{equation}
    \mathcal{L} = \min_{E, G} \ \mathcal{L}_{GAN} + \lambda_M \mathcal{L}_{MSE} + \lambda_p \mathcal{L}_{perceptual} + \lambda_S \mathcal{L}_{SAE}
\end{equation}

where,
\begin{align*}
&\mathcal{L}_{GAN} = E[\log D(x)] + E[\log (1 - D(G(z)))] \\
&\mathcal{L}_{MSE} = \| x - \hat x \|_2 \\
&\mathcal{L}_{perceptual} = \| conv_4(x) - conv_4(\hat x) \|_2 \end{align*}

\section{Experiments}

\subsection{Architecture, Losses, and Hyperparameters}

The network architecture for our encoder and decoder/generator is based on the global generator network proposed by Wang et al. \cite{Wang}, in turn, based on the architecture proposed by Johnson et al.\cite{JohnsonAL16}

\subsubsection*{Encoder}
Let $\texttt{c7s1-k}$ denote a $7 \times 7$ Convolution-Instance Norm-ReLU layer with \textit{k} filters and stride 1. \texttt{dk} denotes a $3 \times 3$ Convolution-Instance Norm-ReLU layer with \textit{k} filters, and stride 2. We use reflection padding to reduce boundary artifacts. \texttt{Rk} denotes a residual block that contains two $3 \times 3$
convolutional layers with the same number of filters on both
layers. \texttt{uk} denotes a $3 \times 3$ fractional-strided-Convolution-
Instance Norm-ReLU layer with \textit{k} filters, and stride $\frac{1}{2}$.

\textbf{Architecture}: \texttt{c7s1-60, d120, d240, d480, d960}

\subsubsection*{Decoder}

Let $\texttt{c3s1-960}$ denote a $3 \times 3$ Convolution-Instance Norm-ReLU layer with 960 filters and stride 1. \texttt{Rk} denotes a residual block that contains two $3 \times 3$
convolutional layers with the same number of filters on both
layers. \texttt{uk} denotes a $3 \times 3$ fractional-strided-Convolution-
Instance Norm-ReLU layer with \textit{k} filters, and stride $\frac{1}{2}$.

\textbf{Architecture}: \texttt{c3s1-960, R960 X 9, u480, u240, u120, u60, c7s1-3}

\begin{figure*} \centering 
\begin{subfigure}[b]{0.3 \linewidth}
\includegraphics[height=5.75cm]{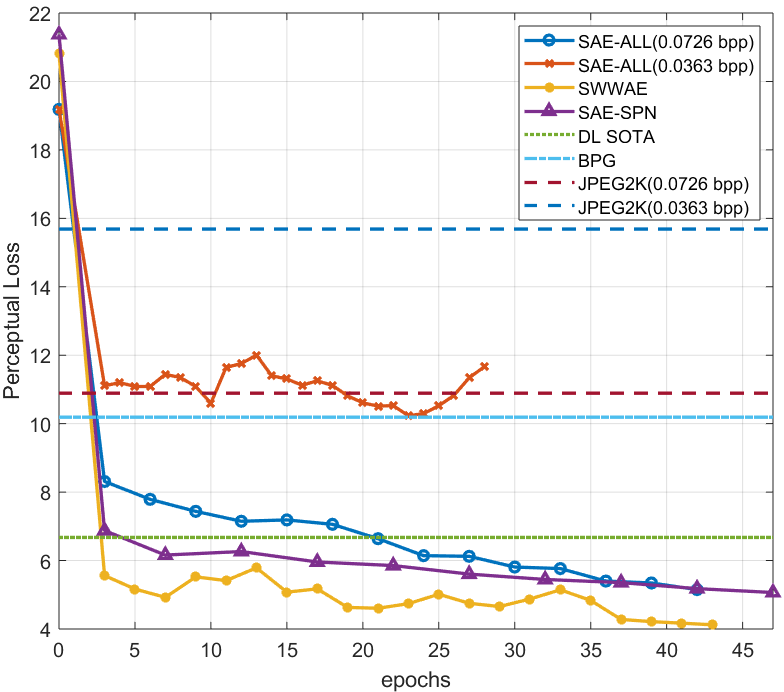}
\caption{Perceptual Loss with training}
\label{fig:7a}
\end{subfigure}\hspace{1 in}
\begin{subfigure}[b]{0.3 \linewidth}
\includegraphics[height=5.75cm]{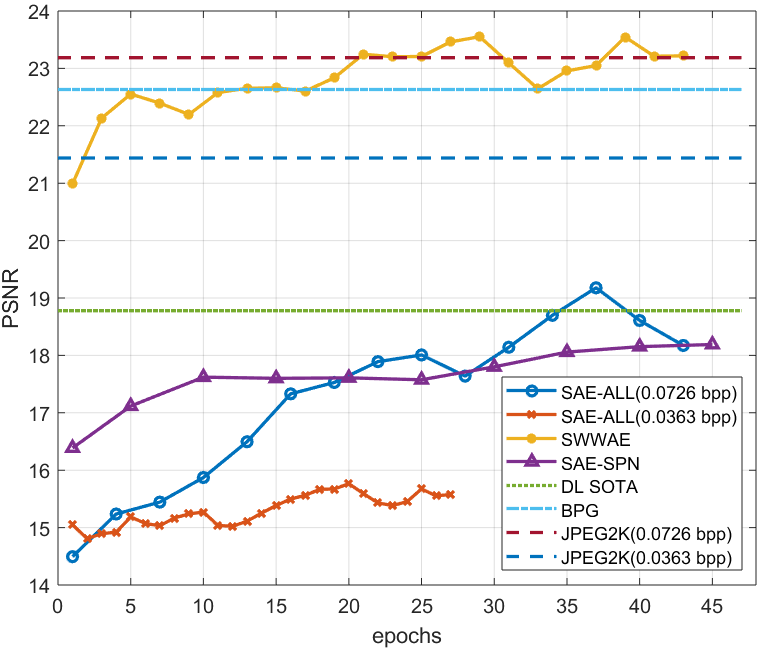}
\caption{PSNR Metric with training}
\label{fig:7b}
\end{subfigure}
\caption{Comparison of Perceptual and Traditional Metrics across models with training}
\label{fig:7}
\end{figure*}

\subsubsection*{Discriminator}

Let $\texttt{c4s2p1-k}$ denote a $4 \times 4$ Convolution-Leaky ReLU layer with k filters and stride value as 2 and padding value as 1 with k filters.

\textbf{Architecture}:\texttt{c4s2p1-64, c4s2p1-128, c4s2p1-192, c4s2p1-256, c4s2p1-512, c4s1p1-1}

We have also included a hard quantizer (non-differentiable), with L = 5 centers, $\mathbb{C}$ = \{-2,-1,0,1,2\}, to control the bitrate given by the expression, (Eq. \ref{bpp}). Additionally, we have incorporated sub-pixel convolutions with ICNR initialization \cite{ICNR}, in place of the originally proposed convolution $+$ upsampling in the decoder to get rid of checkerboard artifacts.

The encoder takes in an image of size H x W x 3 and returns a latent space dimension of H/16 x W/16 x $\mathbb{C}$. Hence the operating point characterized by bpp (Eq. \ref{bpp}) is directly related to the parameter $\mathbb{C}$. We experimented with the performance of our models, at C = \{4,8\} corresponding to 0.0363 bpp and 0.0726 bpp.

The encoder and the decoder/generator are trained with the Adam optimizer with a learning rate of 2e-3, coupled with a Learning Rate(LR) scheduler with a decay parameter of 0.5 for improved training. The discriminator is trained using the SGD optimizer with a learning rate of 2e-5. 

We are also predicting only the first layer switches for the encoder which is of dimension $256 \times 256 \times 60$, with the rest of the unpooling in the decoder done by Transposed Convolution. The intuition behind predicting the first level switches is while encoding the input onto a latent space, the first pool layer carries the most local information, making it essential to reconstructing the original image.  

\begin{equation}
    \textrm{bpp} = \frac{\textrm{H}/16 \times \textrm{W}/16 \times \textrm{C} \times \log_2{\textrm{L}}}{\textrm{H} \times \textrm{W}}
    \label{bpp}
\end{equation}

To obtain more visually pleasing reconstructions, we adopt $\mathcal{L}_{MSE}$ with a weight, $\lambda_{MSE}$ = 1. Since we look to enhance the perceptual quality of the reconstructions, we incorporate $\mathcal{L}_{perceptual}$ based on AlexNet architecture proposed by \cite{GC} with a weight, $\lambda_{p}$ = 5. In addition to the above losses, we incorporate the vanilla GAN loss $\mathcal{L}_{GAN}$, and the SAE layer loss $\mathcal{L}_{SAE}$ for sharper reconstructions, with weight $\lambda_{S}$ = 1.

\subsection{Datasets and Preprocessing steps}
We train and evaluate our models on the Cityscapes dataset \cite{Cordts2016Cityscapes}. We enhance our models by including the CLIC (Challenge on Learned Image Compression) 2019 dataset to generalize better on the color information. The datasets were augmented to 18000 image patches of size $512 \times 512$ px generated with random crops and flips. Furthermore, Contrast Limited Adaptive Histogram Equalization (CLAHE) \cite{CLAHE} was used to enhance the local contrast of these images before feeding them to the network.

\subsection{Baselines}
We benchmark all of our compression models against traditional and deep learning-based state-of-the-art methods.BPG \cite{BPG} is the current state-of-the-art engineered image compression codec that outperforms the other recent codecs, such as JPEG2000 \cite{JPEG} and WebP\cite{WebP}, in terms of PSNR. Specifically in the extreme-learned compression (bpp $<$ 0.1) setting, generative compression proposed by Agussten \textit{et al.} \cite{abs-1804-02958} is the current deep learning based state-of-the-art method. For evaluation purposes, we use pre-trained weights of the same architecture \cite{JT} for comparison. Apart from the above state-of-the-art methods, we compare our models with other popular and common compression standards such as JPEG2000, operated at similar bitrates, i.e 0.0726 bpp, and 0.0363 bpp.

\subsection{Evaluation Metrics}
We benchmark the performance of all our models with traditional metrics, such as PSNR and SSIM. However, the primary focus is benchmarking based on perceptual quality. 


In that regard, we evaluate the performance against perceptual loss, FSIM$_{c}$, and Fréchet Inception Distance (FID). Perceptual Loss is calculated as the $L^2$ distance between the response of the input and reconstructed image obtained after $4^{th}$ convolutional layer of the AlexNet.FSIM$_{c}$ is a measure that is based on the fact that the human visual system uses low-level features to interpret images. 

A dimensionless quantity called phase congruence is used to calculate the similarity between images.FID is a perceptual quality metric proposed by Heusel \textit{et al.}\cite{FID} specifically for evaluating the GAN synthesized images.FID uses the output features after the third pool layer of an inception \cite{inception} network, modeled using a multivariate Gaussian with mean $\mu$ and $\Sigma$. FID between the input dataset x and reconstructed dataset g, is computed as:
\begin{equation}
\textrm{FID}(x,g) = || \mu_x - \mu_g||_2^2 + \textrm{Tr}(\Sigma_x + \Sigma_g - 2(\Sigma_x\Sigma_g)^{\frac{1}{2}})    
\end{equation}

where FID is a measure of the generated samples' accuracy for approximating the real data distribution. Lower FID values signify that the distance between the real and the generated data distribution is smaller and hence correlates with more accurate image quality and diversity. 

\begin{figure}[h]
\centering
\includegraphics[height=6.5cm]{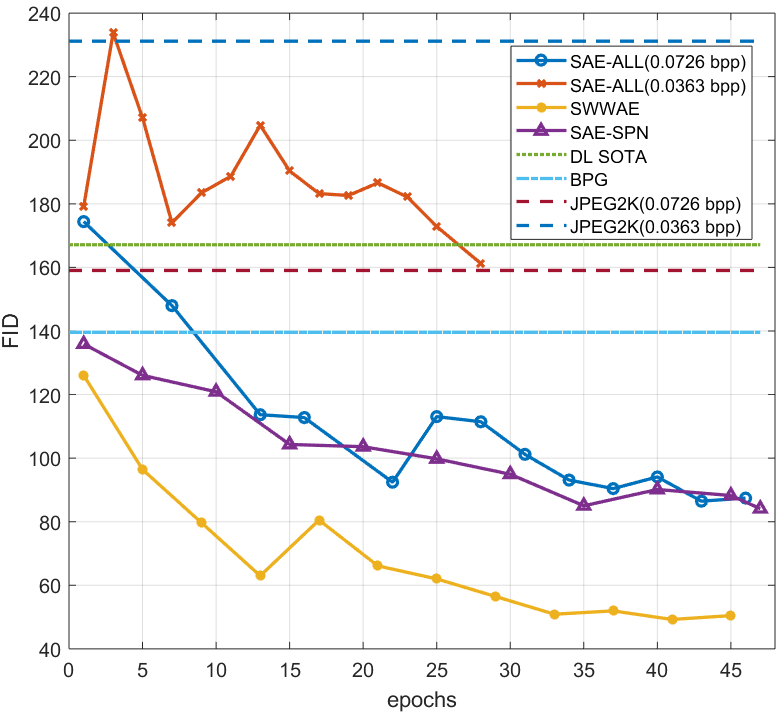}
\caption{Comparison of FID across models with training.}
\label{fig:8}
\end{figure}\hspace{1 in}

\section{Results}

\begin{table*}[h]

\resizebox{\linewidth}{!}{
\begin{tabular}{cc|c|c|c|c|c|}
\cline{2-7}
\multirow{9}{*}{\includegraphics[height = 1.8 in]{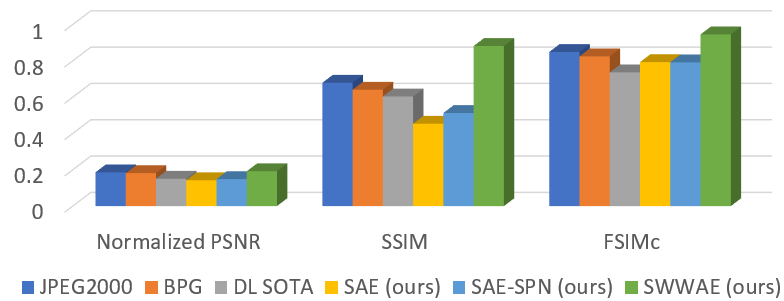}
}

&\multicolumn{1}{|c|}{} &\multicolumn{5}{c|}{Bit Rate = \textbf{0.0726 bpp}} \\ \cline{2-7}
   &    \multicolumn{1}{|c|}{}     & \textbf{SSIM}   & \textbf{PSNR}    & \textbf{FSIM$_c$} & \textbf{PLoss} & \textbf{FID}\\ \cline{2-7}
& \multicolumn{1}{|c|}{\textbf{JPEG2K}\cite{JPEG}}       & 0.6793 & 23.1865 & 0.8491 & 10.89 & 159.05\\ \cline{2-7} 
& \multicolumn{1}{|c|}{\textbf{BPG}\cite{BPG}}        & 0.6411 & 22.6323 & 0.8240 & 10.19 & 139.58\\ \cline{2-7}
& \multicolumn{1}{|c|}{\textbf{DL SOTA}\cite{JT}}     & 0.6035 & 18.7794 & 0.7367 & 6.67 & 167.13\\  \cline{2-7}
& \multicolumn{1}{|c|}{\textbf{SAE (ours) }} & 0.4536 & 17.7478 & 0.7932 & 5.15 & 87.44\\ \cline{2-7}
& \multicolumn{1}{|c|}{\textbf{SAE-SPN (ours)}} & 0.5128 & 18.5084 & 0.7919 & 5.07 & 74.06\\ \cline{2-7}
& \multicolumn{1}{|c|}{\textbf{SWWAE (ours)}} & \textbf{0.8118} & \textbf{23.9258} & \textbf{0.9457} & \textbf{4.17} & \textbf{50.47}\\ \cline{2-7}
& \multicolumn{1}{|c|}{}&\multicolumn{5}{c|}{\textbf{Bit Rate = 0.0363 bpp}} \\ \cline{2-7}
& \multicolumn{1}{|c|}{\textbf{JPEG2K}}     & 0.6002 & 21.4389 & 0.7863 & 15.68 & 231.15\\ \cline{2-7}
& \multicolumn{1}{|c|}{\textbf{SAE (ours)}}  & \textbf{0.3446} & \textbf{15.5972} & \textbf{0.6926} & \textbf{10.19} & \textbf{161.24}\\
\cline{2-7}
(a) Comparison of traditional metrics & \multicolumn{6}{c}{(b) Comparison of perceptual metrics.}\\

\end{tabular}
}
\label{tab1}
\captionof{figure}{Benchmarking our algorithm against competing algorithms}
\end{table*}

To compare the performance across different methods, we plotted the various performance metrics across epochs.

Plots in Fig.\ref{fig:7a} are a representation of the relationship between perceptual loss and epochs for different methods. Following the intuition that since more information on pooling switches is being sent from the encoder to the decoder, the quality of reconstruction achieved with SWWAE is significantly more accurate than its counterparts, like SAE-SPN and SAE-All, and achieves the lowest perceptual loss implying better perceptual quality. SAE-All, CompressNet also shows comparable performance at 0.0726 bpp and far outperforms BPG and JPEG2000.

Plots in Fig.\ref{fig:7b} is a representation of the relationship between PSNR and epochs for different methods. As observed, JPEG 2000 (at 0.0726 bpp) and BPG do appreciably well for PSNR metric, followed by SWWAE, SAE-All (at 0.0726 bpp) and SAE-All (at 0.0363 bpp). This is because traditional compression metrics optimize for PSNR, but fail to visual sharpness, as evident by reconstructions shown in Fig. \ref{fig:ovr}. 

Plots presented in Fig.\ref{fig:8} describes the trends between FID and epochs for different methods. As discussed above, lower FID signifies more accurate approximation to real data distribution and generates visually better looking images. As expected, SWWAE achieves the highest performance in this metric, closely followed by CompressNet performance and SAE-All. This trend follows our intuition of SWWAE and CompressNet performing well on this metric due to the addition of pooling switch information. Traditional compression methods fall behind in this metric, as is evident from the plot. This is because traditional methods optimize for PSNR instead of a perceptual loss.    

The bar plot aptly describes the performance of our methods against BPG and JPEG2000 for different metrics, like Perceptual loss, SSIM and FSIM$_c$. We have benchmarked the performance of our proposed methods against both the traditional and deep learning based state-of-the-art in Fig 8.b. Although perceptual loss and FID are the primary metrics evaluating the visual quality of reconstructions, we have reported results against the traditional metrics as well. The methods we have proposed do comparably well on the traditional metrics and vastly outperform in terms of optimizing for perceptual quality of the image.

\textbf{User Study} : To confirm if the perceptual quality and the FID metric are in accordance with the human perception, we conducted a small scale user study. In the survey, the original image was shown along with the
reconstructed images obtained by three different methods: CompressNet, BPG and JPEG2K. One hundred users from diverse backgrounds were asked to indicate their preference for each pair of reconstructions in the questionnaire. The percentage of their preferred choice has been reported. This clearly validates that CompressNet outperforms the traditional compression methods with superior perceptual quality.

\begin{figure}[h] \centering
\includegraphics[height=4cm]{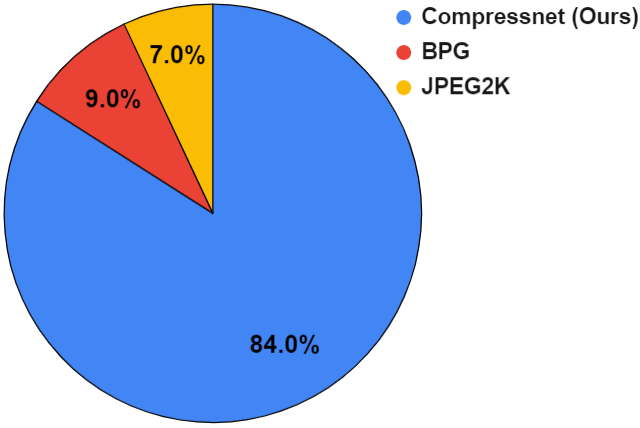}
\caption{ User study results indicating preference on image sharpness and quality across different methods}
\label{fig:7}
\end{figure}

\section{Conclusion}
We have proposed and evaluated different GAN-based frameworks for extreme learned compressions that significantly outperform prior works for extremely low bitrates in terms of visual quality. 
Our proposed model, CompressNet (SAE-SPN) demonstrates quality image compression in results where it performs comparably to traditional methods like JPEG2000 and BPG in terms of PSNR and FSIM$_c$, but is much superior in terms of perceptual quality and FID. We believe learning compressed representations is a promising avenue to learn high-resolution generative models for multimodal data compression as well as adaptive image compression with wide ranging applications.

{\small
\bibliographystyle{ieee}
\bibliography{egpaper}
}

\end{document}